\documentclass{article}

\usepackage{PRIMEarxiv}

\usepackage{amsmath}
\usepackage{amssymb}
\usepackage{amsthm}
\usepackage{algorithm}
\usepackage[noend]{algpseudocode}
\usepackage{booktabs}
\usepackage{epsfig}
\usepackage{graphicx}
\usepackage{epstopdf}
\usepackage{verbatim}
\usepackage{centernot}

\newcommand{\PMCI}{\mathcal{PMC}}

\DeclareMathOperator{\CFL}{CFL}
\DeclareMathOperator{\ICFL}{ICFL}

\newtheorem{theorem}{Theorem}
\newtheorem{definition}{Definition}
\newtheorem{lemma}{Lemma}
\newtheorem{example}{Example}
\newtheorem{proposition}{Proposition}
\newtheorem{corollary}{Corollary}
\newtheorem{remark}{Remark}

\numberwithin{theorem}{section}
\numberwithin{equation}{section}

\title{Unveiling the connection between the Lyndon factorization and the canonical inverse Lyndon factorization
via a border property}

\author{
  Paola Bonizzoni, Brian Riccardi \\
  Dip. di Informatica, Sistemistica e Comunicazione \\
  University of Milano-Bicocca \\
  viale Sarca 336, 20126 Milan, Italy\\
  \texttt{\{paola.bonizzoni,brian.riccardi@unimib.it\}unimib.it} \\
   \And
  Clelia De Felice, Rocco Zaccagnino, Rosalba Zizza \\
  Dip.~di Informatica \\
  University of Salerno \\
  via Giovanni Paolo II 132, 84084 Fisciano, Italy\\
  \texttt{\{cdefelice,rzaccagnino,rzizza\}email@email} \\
}

\begin{document}
\maketitle

\begin{abstract}
The notion of Lyndon word and Lyndon factorization has shown to have unexpected applications in theory as well in developing novel algorithms on words.
A counterpart to these notions are those of inverse Lyndon word and inverse Lyndon factorization. Differently from the Lyndon words,
the inverse Lyndon words may be bordered. The relationship between the two factorizations is related to the inverse lexicographic ordering, and has only been recently explored.
More precisely, a main open question is how to get an inverse Lyndon factorization from a classical Lyndon factorization under the inverse lexicographic ordering, named $\CFL_{in}$.
In this paper we reveal a strong connection between these two factorizations  where the border plays a relevant role. More precisely, we show two main results.
We say that a factorization has the border property if a nonempty border of a factor cannot be a prefix of the next factor.
First we show that there exists a unique inverse Lyndon factorization having the border property.
Then we show that this unique factorization with the border property is the so-called canonical inverse Lyndon factorization, named $\ICFL$.
By showing that $\ICFL$ is obtained by compacting factors of the Lyndon factorization over the inverse lexicographic ordering, we provide a linear time
algorithm for computing $\ICFL$ from $\CFL_{in}$.
\end{abstract}

% keywords can be removed

%%%%%%%%%%%%%%%%%%%%%%%%%%%%%%%%%%%%%%%%%%%%%%%%%%%%%%%%%%%%%%%%
\section{Introduction} \label{intro}

The theoretical investigation of combinatorial properties of well-known word factorizations is a research topic that recently have witnessed special interest especially for improving the efficiency of algorithms. Among these, undoubtedly stands out the \textit{Lyndon Factorization} introduced by Chen, Fox, Lyndon in~\cite{Lyndon}, named $\CFL$. Any word $w$ admits a unique factorization $\CFL(w)$, that is a lexicographically non-increasing sequence of factors which are \textit{Lyndon words}. A Lyndon word $w$ is strictly lexicographically smaller than each of its proper cyclic shifts, or, equivalently, than each of its nonempty proper suffixes~\cite{Lyndon0}. Interesting applications of the use of the Lyndon factorization and Lyndon words are
the development of the bijective Burrows-Wheeler Transforms~\cite{bannai2024constructing,biagi2023number,inplace-bijective-bwt2020} and a novel algorithm for sorting suffixes \cite{lyndonAcc}. In particular,
the notion of a Lyndon word has been re-discovered various times as a theoretical tool to locate short motifs~\cite{delgrange2004star} and relevant k-mers in bioinformatics applications \cite{Martayan2024}. In this line of research,  Lyndon-based word factorizations have been explored to define a novel feature representation for biological sequences based on theoretical combinatorial properties proved to capture sequence similarities~\cite{IS22}.

The notion of a \textit{Lyndon word} has a counterpart that is the notion of an \textit{inverse Lyndon word}, i.e., a word lexicographically greater than its suffixes.
Inverting the relation between a word and its suffixes, as between Lyndon words and inverse Lyndon words, leads to different properties. Indeed, although a word could admit more
than one {\it inverse Lyndon factorization}, that is a factorization into a nonincreasing product of \textit{inverse Lyndon words}, in~\cite{inverseLyndon} the \textit{Canonical Inverse Lyndon Factorization},
named $\ICFL$, was introduced. $\ICFL$ maintains the main properties of $\CFL$: it is unique and can be computed in linear time.
In addition, it maintains a similar {\em Compatibility Property}, used for obtaining the sorting of the suffixes of $w$ (``global suffixes'') by using the sorting of the suffixes of each factor of $\CFL(w)$ (``local suffixes'')~\cite{restivo-sorting-2014}.
Most notably, $\ICFL(w)$ has another interesting property~\cite{inverseLyndon,lata2020,inverseLyndonTCS21}: we can provide an upper bound on the length of the longest common prefix
of two substrings of a word $w$ starting from different positions.

A relationship between $\ICFL(w)$ and $\CFL(w)$ has been proved by using the notion of \textit{grouping}~\cite{inverseLyndon}. First, let $\CFL_{in}(w)$ be the Lyndon factorization
of $w$ with respect to the inverse lexicographic order, it is proved that $\ICFL(w)$ is obtained by concatenating the factors of a non-increasing maximal chain with respect to the prefix order, denoted by $\PMCI$, in $\CFL_{in}(w)$ (see Section \ref{CS}).
Despite this result, the connection between $\CFL_{in}(w)$ and the inverse Lyndon factorization still remained obscure, mainly by the fact that a word may have multiple inverse Lyndon factorizations.

In this paper, we explore this connection between $\CFL_{in}$ and the inverse Lyndon factorizations. Our first main contribution consists in showing that there is a unique inverse Lyndon factorization of a word that has \textit{border property}. The border property states that any nonempty border of a factor cannot be a prefix of the next factor. We further highlight the aforementioned connection by proving that the inverse Lyndon factorization with the border property is a \textit{compact} factorization (Definition \ref{CompFactorization}), i.e., each inverse Lyndon factor is the concatenation of \textit{compact factors}, each obtained by concatenating the longest sequence of identical words in a $\PMCI$. We then show the second contribution of this paper: this unique factorization is $\ICFL$ itself and then provide a simpler linear time algorithm for computing $\ICFL$.
Our algorithm is based on a new property that characterizes $\ICFL(w)$: the last factor in an inverse Lyndon factorization with the border property of $w$ is the longest suffix
of $w$ that is an inverse Lyndon word. Recall that the Lyndon factorization of $w$ has a similar property: the last factor is the longest suffix of $w$ that is a Lyndon word.

%%%%%%%%%%%%%%%%%%%%%%%%%%%%%%%%%%%%%%%%%%%%%%%%%%%%%%%%%%%%%%%%

\section{Words} \label{W}

Throughout this paper we follow
\cite{bpr,CK,Lo,lothaire,reu} for the notations.
We denote by $\Sigma^{*}$ the {\it free monoid}
generated by a finite alphabet $\Sigma$
and we set $\Sigma^+=\Sigma^{*} \setminus 1$, where $1$ is
the empty word.
For a word $w \in \Sigma^*$, we denote by $|w|$ its {\it length}.
A word $x \in \Sigma^*$ is a {\it factor} of $w \in \Sigma^*$ if there are
$u_1,u_2 \in \Sigma^*$ such that $w=u_1xu_2$.
If $u_1 = 1$ (resp. $u_2 = 1$), then $x$ is a {\it prefix}
(resp. {\it suffix}) of $w$.
A factor (resp. prefix, suffix) $x$ of $w$
is {\it proper} if $x \not = w$.
Two words $x,y$ are {\it incomparable} for the prefix order, denoted as $x \Join y$,
if neither $x$ is a prefix of $y$ nor $y$ is a prefix of $x$.
Otherwise, $x,y$ are {\it comparable} for the prefix order.
We write $x \leq_{p} y$ if $x$ is a prefix of $y$
and  $x \geq_{p} y$ if $y$ is a prefix of $x$.
The notion of a pair of words comparable (or incomparable) for the suffix order
is defined symmetrically.

We recall that, given a nonempty word $w$,
a {\it border} of $w$ is a word which is both a proper prefix and a suffix
of $w$ \cite{CHL07}. The longest proper prefix of $w$ which is a suffix of
$w$ is also called {\it the border} of $w$ \cite{CHL07, lothaire}.
A word $w \in \Sigma^+$ is {\it bordered} if it has a nonempty border.
Otherwise, $w$ is {\it unbordered}.
A nonempty word $w$ is \textit{primitive} if
$w = x^k$ implies $k = 1$. An unbordered word is primitive.
A {\it sesquipower} of a word $x$ is a word $w = x^np$ where
$p$ is a proper prefix of $x$ and $n \geq 1$.
Two words $x,y$ are called {\em conjugate} if there exist words
$u,v$ such that $x=uv, y=vu$.
The conjugacy relation is an equivalence relation. A conjugacy class
is a class of this equivalence relation.

\begin{definition} \label{lex-order}
Let $(\Sigma, <)$ be a totally ordered alphabet.
The {\it lexicographic} (or {\it alphabetic order})
$\prec$ on $(\Sigma^*, <)$ is defined by setting $x \prec y$ if
\begin{itemize}
\item $x$ is a proper prefix of $y$,
or
\item $x = ras$, $y =rbt$, $a < b$, for $a,b \in \Sigma$ and $r,s,t \in \Sigma^*$.
\end{itemize}
\end{definition}

In the next part of the paper we will implicitly refer to
totally ordered alphabets.
For two nonempty words $x,y$, we write $x \ll y$ if
$x \prec y$ and $x$ is not a proper prefix of $y$
\cite{Bannai15}. We also write $y \succ x$ if $x \prec y$.
Basic properties of the lexicographic order are recalled below.

\begin{lemma} \label{proplexord}
For $x,y \in \Sigma^+$, the following properties hold.
\begin{itemize}
\item[(1)]
$x \prec y$ if and only if $zx \prec zy$,
for every word $z$.
\item[(2)]
If $x \ll y$, then $xu \ll yv$
for all words $u,v$.
\item[(3)]
If $x \prec y \prec xz$ for a word $z$,
then $y = xy'$ for some word $y'$ such that $y' \prec z$.
\item[(4)]
If $x \ll y$ and $y \ll z$, then $x \ll z$.
\end{itemize}
\end{lemma}

Let $\mathcal{S}_1, \ldots , \mathcal{S}_t$ be sequences, with
$\mathcal{S}_j = (s_{j,1}, \ldots , s_{j,r_j})$.
For abbreviation, we let
$(\mathcal{S}_1, \ldots , \mathcal{S}_t)$
stand for the sequence
$(s_{1,1}, \ldots , s_{1,r_1}, \ldots , s_{t,1}, \ldots , s_{t,r_t})$.

%%%%%%%%%%%%%%%%%%%%%%%%%%%%%%%%%%%%%%%%%%%%%%%%%%%%%%%%%%%%%%%%
\section{Lyndon words} \label{LW}

\begin{definition}\label{Lyndon-word}
A Lyndon word $w \in \Sigma^+$ is a word which is primitive and the smallest one
in its conjugacy class for the lexicographic order.
\end{definition}

\begin{example} \label{ppp}
Let $\Sigma = \{a,b\}$ with $a < b$.
The words $a$, $b$, $aaab$, $abbb$, $aabab$ and $aababaabb$
are Lyndon words. On the contrary, $abab$, $aba$ and
$abaab$ are not Lyndon words.
\end{example}

\begin{proposition} \label{Pr1}
Each Lyndon word $w$ is unbordered.
\end{proposition}

A class of conjugacy is also called a {\it necklace} and often identified
with the minimal word for the lexicographic
order in it. We will adopt this terminology. Then
a word is a necklace if and only if it is a power of a Lyndon word.
A {\it prenecklace} is a prefix of a necklace. Then clearly
any nonempty prenecklace $w$ has the form $w = (uv)^ku$, where $uv$ is a Lyndon word,
$u \in \Sigma^*$, $v \in \Sigma^+$, $k \geq 1$, that is, $w$ is a sesquipower of a Lyndon word $uv$.
The following result has been proved in \cite{duval}.
It shows that the nonempty prefixes of Lyndon words are exactly
the nonempty prefixes of the powers of Lyndon words with the exclusion of
$c^k$, where $c$ is the maximal letter and $k \geq 2$.

\begin{proposition} \label{preprime}
A word is a nonempty prefix of a Lyndon word if and only if it is a sesquipower of a Lyndon word distinct of
$c^k$, where $c$ is the maximal letter and $k \geq 2$.
\end{proposition}

In the following $L = L_{(\Sigma^*, <)}$ will be the set of Lyndon words, totally
ordered by the relation $\prec$ on $(\Sigma^*, <)$.

\begin{theorem} \label{Lyndon-factorization}
Any word $w \in \Sigma^+$ can be written in a unique way as
a nonincreasing product $w=\ell_1 \ell_2 \cdots \ell_h$ of Lyndon words, i.e., in the form
\begin{eqnarray} \label{LF}
w & = & \ell_1 \ell_2 \cdots \ell_h, \mbox{ with } \ell_j \in L \mbox{ and } \ell_1 \succeq \ell_2 \succeq \ldots \succeq \ell_h
\end{eqnarray}
\end{theorem}

The sequence $\CFL(w) = (\ell_1, \ldots, \ell_h)$ in Eq. \eqref{LF} is called the
{\it Lyndon decomposition} (or {\it Lyndon factorization}) of $w$.
It is denoted by $\CFL(w)$ because Theorem \ref{Lyndon-factorization}
is usually credited to Chen, Fox and Lyndon
\cite{Lyndon}.
The following result, proved
in \cite{duval}, is necessary for our aims.

\begin{corollary} \label{cor1}
Let $w \in \Sigma^+$, let $\ell_1$ be its longest
prefix which is a Lyndon word and let $w'$ be
such that $w= \ell_1 w'$. If $w' \not = 1$, then $\CFL(w) = (\ell_1, \CFL(w'))$.
\end{corollary}

Sometimes we need to emphasize consecutive equal factors
in $\CFL$. We write
$\CFL(w) = (\ell_1^{n_1}, \ldots, \ell_r^{n_r})$ to denote
a tuple of $n_1 + \ldots + n_r$ Lyndon words,
where $r  > 0$, $n_1, \ldots , n_r \geq 1$. Precisely
$\ell_1 \succ \ldots \succ \ell_r$
are Lyndon words, also named {\it Lyndon factors} of $w$.
There is a linear time algorithm
to compute the pair $(\ell_1, n_1)$ and thus, by iteration,
the Lyndon factorization of $w$ \cite{FM,lothaire}.
Linear time algorithms may also be found in \cite{duval}
and in the more recent paper
\cite{GGT}.

%%%%%%%%%%%%%%%%%%%%%%%%%%%%%%%%%%%%%%%%%%%%%%%%%%%%%%%%%%%%%%%
\section{Inverse Lyndon words} \label{ILW}

For the material in this section see \cite{inverseLyndon,lata2020,inverseLyndonTCS21}.
Inverse Lyndon words are related to the
inverse alphabetic order. Its definition is recalled below.

\begin{definition} \label{ILO}
Let $(\Sigma, <)$ be a totally ordered alphabet.
The {\rm inverse} $<_{in}$ of $<$ is defined
by
$$ \forall a, b \in \Sigma \quad b <_{in} a \Leftrightarrow a < b $$
The {\rm inverse lexicographic} or {\rm inverse alphabetic order}
on $(\Sigma^*, <)$, denoted $\prec_{in}$, is the lexicographic order
on $(\Sigma^*, <_{in})$.
\end{definition}

\begin{example}
Let $\Sigma = \{a,b,c,d \}$ with $a < b < c < d$. Then $dab \prec dabd$
and $dabda \prec dac$. We have
$d <_{in} c <_{in} b <_{in} a$. Therefore $dab \prec_{in} dabd$ and $dac \prec_{in} dabda$.
\end{example}

Of course for all $x, y \in \Sigma^*$ such that $x \Join y$,
$$y \prec_{in} x \Leftrightarrow x \prec y .$$
Moreover, in this case $x \ll y$. This justifies
the adopted terminology.

From now on, $L_{in} = L_{(\Sigma^*, <_{in})}$ denotes the set
of the Lyndon words on $\Sigma^*$ with respect to the inverse lexicographic order.
Following \cite{antiLyndon},
a word $w \in L_{in}$ will be named an {\it anti-Lyndon word}. Correspondingly, an
{\it anti-prenecklace} will be a prefix of an {\it anti-necklace}, which in turn will
be a necklace with respect to the inverse lexicographic order.

In the following, we denote by $\CFL_{in}(w)$ the Lyndon factorization
of $w$ with respect to the inverse order $<_{in}$.

\begin{definition} \label{inverse-Lyndon-word}
A word $w \in \Sigma^+$ is an inverse Lyndon word if
$s \prec w$, for each nonempty proper suffix $s$ of $w$.
\end{definition}

\begin{example}
{\rm The words $a$, $b$, $aaaaa$, $bbba$, $baaab$, $bbaba$ and $bbababbaa$
are inverse Lyndon words on $\{a,b\}$, with $a < b$.
On the contrary, $aaba$ is not an inverse Lyndon word since $aaba \prec ba$.
Analogously, $aabba \prec ba$ and thus $aabba$ is not an inverse Lyndon word.}
\end{example}

The following result, proved in \cite{inverseLyndon,inverseLyndonTCS21},
and also in \cite{IL2023}, summarizes some properties of the inverse Lyndon words.

\begin{proposition} \label{varie}
Let $w \in \Sigma^+$. Then we have
\begin{enumerate}
\item
The word $w$ is an anti-Lyndon word if and only if it is
an unbordered inverse Lyndon word.
\item
The word $w$ is an inverse Lyndon word if and only if
$w$ is a nonempty anti-prenecklace.
\item
If $w$ is an inverse Lyndon word, then any nonempty prefix of $w$
is an inverse Lyndon word.
\end{enumerate}
\end{proposition}

\begin{definition} \label{inverse factorization}
An inverse Lyndon factorization of
a word $w \in \Sigma^+$ is a sequence $(m_1, \ldots, m_{k})$ of inverse Lyndon words
such that $m_1 \cdots m_{k} = w$ and $m_i \ll m_{i+1}$, $1 \leq i \leq k-1$.
\end{definition}

As the following example in \cite{inverseLyndon}
shows, a word may have different
inverse Lyndon factorizations.

\begin{example} \label{nonunique}
{\rm Let $\Sigma = \{a,b,c,d \}$ with $a < b < c < d$,
$z = dabdadacddbdc$.
It is easy to see that
$(dab,dadacd,db,dc)$, $(dabda,dac,ddbdc)$, $(dab, dadac, ddbdc)$
are all inverse Lyndon factorizations of
$z$.}
\end{example}

%%%%%%%%%%%%%%%%%%%%%%%%%%%%%%%%%%%%%%%%%%%%%%%%%%%%

\section{The border property} \label{SectionBorder}

In this section we prove the main result of this paper, namely, for any nonempty word $w$,
there exists a unique inverse Lyndon factorization of $w$ which has
a special property, named the border property.

\begin{definition} [Border property] \label{BorderFactorization}
Let $w \in \Sigma^+$.
A factorization $(m_1, \ldots , m_k)$
of $w$ {\rm has the border property} if each nonempty border
$z$ of $m_i$ is not a prefix of $m_{i+1}$, $1 \leq i \leq k-1$.
\end{definition}

We first prove a fundamental property of the
inverse Lyndon factorizations of $w$ which have the border property.

\begin{lemma} \label{fundamental}
Let $w \in \Sigma^+$, let $(m_1, \ldots , m_k)$ be an
inverse Lyndon factorization of $w$ having the border property.
If $\alpha$ is a nonempty border of $m_j$, $1 \leq j \leq k-1$, then
there exists a nonempty prefix $\beta$ of $m_{j + 1}$
such that $|\beta| \leq |\alpha|$ and $\alpha \ll \beta$.
\end{lemma}
\begin{proof}
Let $w \in \Sigma^+$, let $(m_1, \ldots , m_k)$ be an
inverse Lyndon factorization of $w$ having the border property, let
$\alpha$ be a nonempty border of $m_j$, $1 \leq j \leq k-1$.
We distinguish two cases: either $|m_{j + 1}| < |\alpha|$ or
$|m_{j + 1}| \geq |\alpha|$.

Assume $|m_{j + 1}| < |\alpha|$. By hypothesis $(m_1, \ldots , m_k)$ is an
inverse Lyndon factorization, hence $m_j \ll m_{j + 1}$, that is, there are
$r,s,t \in \Sigma^*$, $a,b \in \Sigma$, such that $a < b$ and
$m_j = ras$, $m_{j + 1} = rbt$. Obviously $|ra| \leq |m_{j + 1}| < |\alpha|$, thus
there is $s' \in \Sigma^*$ such that $\alpha = ra s'$. Consequently,
$\alpha = ra s' \ll rbt = m_{j + 1}$ and our claim holds with
$\beta = m_{j + 1}$.

Assume $|m_{j + 1}| \geq |\alpha|$. Let $\beta$ be the nonempty prefix
of $m_{j + 1}$ such that $|\beta| = |\alpha|$.
Clearly $\beta \not = \alpha$ because $(m_1, \ldots , m_k)$
has the border property.
Since $\alpha$ and $\beta$ are two different nonempty words of the same length,
either $\beta \ll \alpha$ or $\alpha \ll \beta$.
The first case leads to a contradiction because if $\beta \ll \alpha$ then
$m_{j + 1} \ll m_j$ by Lemma \ref{proplexord} and this contradicts the fact that
$(m_1, \ldots , m_k)$ is an inverse Lyndon factorization. Thus, $\alpha \ll \beta$
and the proof is complete.
\end{proof}

\begin{proposition} \label{UnicaBorder}
For each $w \in \Sigma^+$,
there exists a unique inverse Lyndon factorization of $w$
having the border property.
\end{proposition}
\begin{proof}
The proof is by induction on $|w|$. If $|w| = 1$, then $F_1(w) = F_2(w) = (w)$
and statement clearly holds. Thus assume $|w| > 1$.
Let $F_1(w) = (f_1, \ldots, f_k)$ and $F_2(w) = (f'_1, \ldots, f'_v)$
two inverse Lyndon factorizations of $w$ having the border property.
Thus
\begin{equation} \label{EQ0}
f_1 \cdots f_k = f'_1 \cdots f'_v = w
\end{equation}
If $|f_k| = |f'_v|$ and $v = 1$ or $k = 1$, clearly $f_k = f'_v$
and $F_1(w) = F_2(w)$. Analogously, if $|f_k| = |f'_v|$,
$v > 1$ and $k > 1$, then $f_k = f'_v$ and
$F'_1(w') = (f_1, \ldots, f_{k - 1})$, $F'_2(w') = (f'_1, \ldots, f'_{v - 1})$ would be
two inverse Lyndon factorizations of $w'$ having the border property, where
$w'$ is such that $w = w' f_k$. Of course, $|w'| < |w|$. By induction hypothesis,
$F'_1(w') = F'_2(w')$, hence $F_1(w) = F_2(w)$.

By contradiction, let $|f_k| \not = |f'_v|$. Assume $|f_k| < |f'_v|$
(similar arguments apply if $|f_k| > |f'_v|$).
The word $f_k$ is a proper suffix of $f'_v$. Clearly $k > 1$.
Let $g$ be the smallest integer such that
$f_{g+1} \cdots  f_k$ is a proper suffix of $f'_v$,
$1 \leq g \leq k - 1$, that is,
\begin{equation} \label{EQ1}
f'_v = \alpha f_{g+1} \cdots f_k
\end{equation}
where $\alpha \in \Sigma^+$ is a suffix of $f_{g}$.

Notice that
\begin{equation} \label{EQ2}
\alpha \not \ll f_{g+1}
\end{equation}
Indeed, if $\alpha \ll f_{g+1}$, then, by Eq. \eqref{EQ1}, we would have
$f'_v = \alpha f_{g+1} \cdots f_k \ll f_{g+1} \cdots f_k$,
which is impossible because $f'_v$ is an inverse Lyndon word.

The word $\alpha$ is a nonempty proper suffix of $f_{g}$
since otherwise we would have
$\alpha = f_g \ll f_{g+1}$, contrary to Eq. \eqref{EQ2}.
Since $f_{g}$ is an inverse Lyndon word and
$\alpha$ is a nonempty proper suffix of $f_{g}$, either
$\alpha \leq_p f_{g}$ or $\alpha \ll f_{g}$.

If $\alpha \leq_p f_{g}$, then $\alpha$ is a nonempty border of $f_{g}$,
then, by Lemma \ref{fundamental}, there exists a nonempty prefix $\beta$ of $f_{g+1}$
such that $|\beta| \leq |\alpha|$ and $\alpha \ll \beta$.
Thus, $\alpha \ll f_{g+1}$ which contradicts Eq. \eqref{EQ2}.
Assume $\alpha \ll f_{g}$.
Since $f_{g} \ll f_{g+1}$,
by Lemma \ref{proplexord} we have $\alpha \ll f_{g+1}$ which contradicts
once again Eq. \eqref{EQ2}.
This finishes the proof.
\end{proof}

%%%%%%%%%%%%%%%%%%%%%%%%%%%%%%%%%%%%%%%%%%%%%%

\section{Groupings and compact factorizations} \label{CS}

In this section we prove a structural property of
an inverse Lyndon factorization having the border property, namely
it is a \emph{compact factorization}.
This result is crucial to characterize the relationship between $\CFL_{in}(w)$
and the factorization into inverse Lyndon words of $w$.
First we report the notion of {\it grouping} given in
\cite{inverseLyndon}.
We refer to \cite{inverseLyndon,inverseLyndonTCS21}
for a detailed and complete discussion on this topic.

Let $\CFL_{in}(w) = (\ell_1, \ldots , \ell_h)$, where
$\ell_1 \succeq_{in} \ell_2 \succeq_{in} \ldots \succeq_{in} \ell_h$.
Consider the partial order $\geq_p$, where
$x \geq_p y$ if $y$ is a prefix of $x$. Recall that a {\it chain} is a set
of a pairwise comparable elements. We say that a chain is maximal if it is
not strictly contained in any other chain.
A non-increasing {\it (maximal) chain}
in $\CFL_{in}(w)$ is the
sequence corresponding to a (maximal) chain in the
multiset $\{\ell_1, \ldots , \ell_h \}$ with respect to $\geq_p$.
We denote by $\PMCI$ a non-increasing maximal chain in $\CFL_{in}(w)$.
Looking at the definition of the (inverse) lexicographic order,
it is easy to see that a $\PMCI$ is a sequence
of consecutive factors in $\CFL_{in}(w)$.
Moreover $\CFL_{in}(w)$ is
the concatenation of its $\PMCI$.
The formal definitions are given below.

\begin{definition} \label{MaxCh}
Let $w \in \Sigma^+$, let $\CFL_{in}(w) = (\ell_1, \ldots , \ell_h)$
and let $1 \leq r < s \leq h$.
We say that $\ell_{r}, \ell_{r+1}, \ldots , \ell_{s}$
is a non-increasing {\rm maximal chain for the prefix order}
in $\CFL_{in}(w)$, abbreviated $\PMCI$, if
$\ell_{r} \geq_p \ell_{r+1} \geq_p \ldots  \geq_p \ell_{s}$.
Moreover, if $r > 1$, then $\ell_{r - 1} \not \geq_p \ell_{r}$,
if $s < h$, then $\ell_{s} \not \geq_p \ell_{s +1}$.
Two $\PMCI$ $\mathcal{C}_1 = \ell_{r}, \ell_{r+1}, \ldots , \ell_{s}$,
$\mathcal{C}_2 = \ell_{r'}, \ell_{r'+1}, \ldots , \ell_{s'}$ are
{\rm consecutive} if $r' = s+1$ (or $r = s' +1$).
\end{definition}

\begin{definition} \label{DecMaxCh}
Let $w \in \Sigma^+$, let $\CFL_{in}(w) = (\ell_1, \ldots , \ell_h)$.
We say that $(\mathcal{C}_{1}, \mathcal{C}_{2}, \ldots , \mathcal{C}_{s})$
is {\rm the decomposition} of $\CFL_{in}(w)$ into its non-increasing maximal chains for the prefix order
if the following holds
\begin{itemize}
\item[(1)]
Each $\mathcal{C}_{j}$ is a non-increasing maximal chain in $\CFL_{in}(w)$.
\item[(2)]
$\mathcal{C}_{j}$ and $\mathcal{C}_{j + 1}$ are consecutive, $1 \leq j \leq s-1$.
\item[(3)]
$\CFL_{in}(w)$ is the concatenation of the sequences $\mathcal{C}_{1}, \mathcal{C}_{2}, \ldots , \mathcal{C}_{s}$.
\end{itemize}
\end{definition}

\begin{definition} \label{groupingNew}
Let $w \in \Sigma^+$.
We say that $(m_1, \ldots , m_k)$ is a {\rm grouping} of $\CFL_{in}(w)$ if
the following holds
\begin{itemize}
\item[(1)]
$(m_1, \ldots , m_k)$ is an inverse Lyndon factorization of $w$
\item[(2)]
Each factor $m_j$, is the product of consecutive factors in
a $\PMCI$ in $\CFL_{in}(w)$.
\end{itemize}
\end{definition}

\begin{example} \label{groupingsesempi}
{\rm Let $\Sigma = \{a,b,c,d \}$, $a < b < c < d$,
and $w = dabadabdabdadac$.
We have $\CFL_{in}(w) = (daba, dab, dab, dadac)$.
The decomposition of $\CFL_{in}(w)$ into its $\PMCI$
is $((daba, dab, dab), (dadac))$. Moreover,
$(daba, dabdab, dadac)$ is a grouping of $\CFL_{in}(w)$ but for the
inverse Lyndon factorization
$(dabadab, dabda, dac)$ this is no longer true.

Next, let $y = dabadabdabdabdadac$. We have
$\CFL_{in}(y) = (daba, dab, dab, dab, dadac)$.
The decomposition of $\CFL_{in}(w)$ into its $\PMCI$
is $((daba, dab, dab, dab), (dadac))$.
Moreover, $(daba, (dab)^3, dadac)$ and $(dabadab, (dab)^2, dadac)$
are two groupings of $\CFL_{in}(y)$.}
\end{example}

For our aims, we need to consider the words that are concatenations
of equal factors in $\CFL_{in}$.
This approach leads to a refinement of the partition of $\CFL_{in}$
into non-increasing maximal chains for the prefix order, as defined below.

\begin{definition}[Compact sequences] \label{def:equal-chain}
Let ${\cal C}=(\ell_1, \ldots , \ell_h)$ be a non-increasing maximal chain for the prefix order
in $\CFL_{in}(w)$. The decomposition of ${\cal C}$ into {\rm maximal compact sequences}
is the sequence $({\cal G}_1, \ldots, {\cal G}_n)$ such that
\begin{itemize}
\item[(1)]
${\cal C} =({\cal G}_1, \ldots, {\cal G}_n)$
\item[(2)]
For every $i$, $1 \leq i \leq n$, ${\cal G}_i$ consists of the longest sequence of consecutive identical elements in ${\cal C}$
\end{itemize}
Let $(\mathcal{C}_{1}, \mathcal{C}_{2}, \ldots , \mathcal{C}_{s})$
be the decomposition of $\CFL_{in}(w)$ into its non-increasing maximal chains for the prefix order.
The decomposition of $\CFL_{in}(w)$ into its maximal compact sequences
is obtained by replacing each $\mathcal{C}_{j}$ in $(\mathcal{C}_{1}, \mathcal{C}_{2}, \ldots , \mathcal{C}_{s})$
with its decomposition into maximal compact sequences.
\end{definition}

\begin{definition}[Compact factor] \label{CompFactor}
Let $({\cal G}_1, \ldots, {\cal G}_n)$ be the decomposition of $\CFL_{in}(w)$ into its maximal compact sequences.
For every $i$, $1 \leq i \leq n$, the concatenation $g_i$ of the elements in ${\cal G}_i$
is a \rm{compact factor} in $\CFL_{in}(w)$.
\end{definition}

\begin{definition} [Compact factorization] \label{CompFactorization}
Let $w \in \Sigma^+$.
We say that $(m_1, \ldots , m_k)$ is a {\rm compact factorization}
of $w$ if $(m_1, \ldots , m_k)$ is an inverse Lyndon factorization of $w$
and each $m_j$, $1 \leq j \leq k$, is a concatenation of compact factors in $\CFL_{in}(w)$.
\end{definition}

\begin{example} \label{compactesempi}
{\rm Consider again $y = dabadabdabdabdadac$
over $\Sigma = \{a,b,c,d \}$, $a < b < c < d$,
as in Example \ref{groupingsesempi}.
The decomposition of $\CFL_{in}(y) = (daba, dab, dab, dab, dadac)$
into its maximal compact sequences is
$((daba), (dab, dab, dab), (dadac))$. The compact factors in $\CFL_{in}(w)$
are $daba, (dab)^3, dadac$.
Moreover, $(daba, (dab)^3, dadac)$ is a compact factorization
whereas $(dabadab, (dab)^2, dadac)$
is a grouping of $\CFL_{in}(y)$ which is not a compact factorization.}
\end{example}

\begin{proposition} \label{BorderCompact}
Let $w \in \Sigma^+$.
If $(m_1, \ldots , m_k)$ is an inverse Lyndon factorization of $w$
having the border property, then
$(m_1, \ldots , m_k)$ is a compact factorization
of $w$.
\end{proposition}
\begin{proof}
Let $w \in \Sigma^+$, let
$(m_1, \ldots , m_k)$ be an inverse Lyndon factorization of $w$
having the border property.
Let $\CFL_{in}(w) = (\ell_1, \ldots , \ell_h)$, where
$\ell_1 \succeq_{in} \ell_2 \succeq_{in} \ldots \succeq_{in} \ell_h$ and
$\ell_1, \ldots , \ell_h$ are anti-Lyndon words.
First we prove that $(m_1, \ldots , m_k)$ is a grouping
of $\CFL_{in}(w)$ by induction on $|w|$. If $|w| = 1$ the statement clearly holds,
thus assume $|w| > 1$.

The words $m_1$ and $\ell_1$ are
comparable for the prefix order, hence either $m_1$ is a proper prefix of $\ell_1$
or $\ell_1$ is a prefix of $m_1$.
Suppose that $m_1$ is a proper prefix of $\ell_1$.
Thus, there are $j$, $1 < j \leq k$, and $x, y \in \Sigma^*$, $x \not = 1$,
such that $m_j = xy$ and $\ell_1 = m_1 \cdots m_{j-1} x$.
Necessarily it turns out $j = 2$ because otherwise
$m_1 \ll  m_{j-1}$, hence, by Lemma \ref{proplexord}, $\ell_1 \ll  m_{j-1}x$
and this contradicts the fact that $\ell_1$ is an anti-Lyndon word.
In conclusion $\ell_1 = m_1x$ and $m_2 = xy$.
We know that $m_1 \ll  m_2$, that is, there are
$r,s,t \in \Sigma^*$, $a,b \in \Sigma$, such that $a < b$ and
$m_1 = ras$, $m_2 = rbt = xy$.
If $|x| \leq |r|$, then $r$ is a nonempty border of $\ell_1$
and if $|x| > |r|$, then there is a word $t'$ such that $x = rbt'$
which implies $\ell_1 \ll x$.
Both cases again contradict the fact that $\ell_1$ is an anti-Lyndon word.

Therefore, $\ell_1$ is a prefix of $m_1$.
Let $i$ be the largest integer such that
$m_1 = \ell_1 \cdots \ell_{i-1} x$,
$x, y \in \Sigma^*$, $\ell_i = xy$, $1 < i \leq h$, $y \not = 1$.
Let $(\mathcal{C}_{1}, \mathcal{C}_{2}, \ldots , \mathcal{C}_{s})$
be the decomposition of $\CFL_{in}(w)$ into its non-increasing maximal chains for the prefix order.
We claim that $\ell_1 \cdots \ell_{i - 1}$ is a prefix of the concatenation of the elements of
$\mathcal{C}_{1}$, thus $(\ell_1, \ldots , \ell_{i - 1})$ is a chain for the prefix order.
If $i = 1$ we are done. Let $ i > 1$.
By contradiction, assume that there is $j$, $1 < j < i$,
such that $\ell_j \not \in \mathcal{C}_{1}$.
Therefore, $\ell_1 \ll \ell_j$ which implies
$m_1 \ll \ell_j \cdots \ell_{i-1} x$
and this contradicts the fact that $m_1$ is an inverse Lyndon word.

We now prove that $x = 1$. Assume $x \not = 1$.
As a preliminary step, we prove that there is no nonempty prefix $\beta$ of $m_{2}$ such that
$|\beta| \leq |x|$ and $x \ll \beta$. In fact, if such a prefix existed, there would be
$r,s,t \in \Sigma^*$, $a,b \in \Sigma$, such that $a < b$ and
$x = ras$, $\beta = rbt$. If $|\ell_i| \leq |x r|$ then $\ell_i  = x r' = rasr'$, where
$r'$ would be a nonempty prefix of $r$, thus a nonempty border of $\ell_i$
(recall that $\ell_i = xy$ with $y \not = 1$).
If $|\ell_i| > |x r|$, then there would be a word $t'$ such that
$\ell_i = rasrbt'$ which would imply $\ell_i \ll rbt'$.
Both cases contradict the fact that $\ell_i$ is an anti-Lyndon word.

If $x \not = 1$, then either $\ell_i$ is a prefix of $\ell_1$
or $\ell_1 \ll \ell_i$.
If $\ell_i$ were a prefix of $\ell_1$, then $x$ would be a nonempty border of $m_1$.
By Lemma \ref{fundamental} there would exist a nonempty prefix $\beta$ of $m_{2}$
such that $|\beta| \leq |x|$ and $x \ll \beta$ which contradicts our preliminary step.

If it were true that $\ell_1 \ll \ell_i$ then there would be
$r,s,t \in \Sigma^*$, $a,b \in \Sigma$, such that $a < b$ and
$\ell_1 = ras$, $\ell_i = rbt = xy$.
If $|x| > |r|$, then there would be a word $t'$ such that $x = rbt'$
which would imply $m_1 \ll x$ and this contradicts the fact that $m_1$ is an inverse Lyndon word.
If $|x| \leq |r|$, then $x$ is a prefix of $r$ and is a nonempty border of $m_1$.
By Lemma \ref{fundamental} again, there would exist a nonempty prefix $\beta$ of $m_{2}$
such that $|\beta| \leq |x|$ and $x \ll \beta$ which contradicts again our preliminary step.

Let $w' \in \Sigma^*$ be such that $w = m_1 w'$. If $w' = 1$ we are done.
Assume $w' \not = 1$. Clearly $|w'| < |w|$.
Of course $(m_2, \ldots , m_k)$ is an inverse Lyndon factorization of $w$
having the border property.
Moreover, by Corollary \ref{cor1},
$\CFL_{in}(w') = (\ell_i, \ldots , \ell_h)$
and $(\mathcal{C}'_{1}, \mathcal{C}_{2}, \ldots , \mathcal{C}_{s})$
is the decomposition of $\CFL_{in}(w')$ into its non-increasing maximal chains for the prefix order,
where $\mathcal{C}'_{1}$ is defined by
$\mathcal{C}_{1} = (\ell_1, \ldots , \ell_{i - 1}, \mathcal{C}'_{1})$.
By induction hypothesis, $(m_2, \ldots , m_k)$ is a grouping
of $\CFL_{in}(w')$ and consequently $(m_1, \ldots , m_k)$ is a grouping
of $\CFL_{in}(w)$.

Finally, to obtain a contradiction, suppose that
$(m_1, \ldots , m_k)$ is a grouping
of $\CFL_{in}(w)$ having the border property
such that $(m_1, \ldots , m_k)$ is not a compact factorization
of $w$. To adapt the notation to the proof, set
$\CFL_{in}(w) = (\ell_1^{n_1}, \ldots, \ell_r^{n_r})$,
where $r  > 0$, $n_1, \ldots , n_r \geq 1$ and
$\ell_1, \ldots , \ell_r$ are anti-Lyndon words.
By Definitions \ref{groupingNew} and \ref{CompFactorization},
there exist integers $j, h, p_h, q_h$, $1 \leq j \leq k - 1$, $1 \leq h \leq r$,
$p_h \geq 1$, $q_h \geq 1$, $p_h + q_h \leq n_h$, such that
$m_j$ ends with $\ell_h^{p_h}$ and $m_{j + 1}$ starts with $\ell_h^{q_h}$.
Thus, by Definition \ref{groupingNew}, $\ell_h$ is a prefix of $m_j$.
Moreover, $\ell_h$ is a proper prefix of $m_j$. Indeed otherwise
$\ell_h = m_j \leq_p m_{j + 1}$ which is impossible because
$m_j \ll m_{j + 1}$ ($(m_1, \ldots , m_k)$ is an inverse Lyndon factorization).
Thus $\ell_h$ is a nonempty border of $m_j$.
The word $\ell_h$ is also a prefix of $m_{j + 1}$
and this contradicts the fact that
$(m_1, \ldots , m_k)$ has the border property.
\end{proof}

%%%%%%%%%%%%%%%%%%%%%%%%%%%%%%%%%%%%%%%%%%%%%%%%%%%%%%%%%%%%%%%%%%%%%
\section{The canonical inverse Lyndon factorization: the algorithm} \label{icfl}

In this section we state another relevant result of the paper related to the main one
stated in Section \ref{SectionBorder}.
We have shown that a nonempty word $w$ can have more than one inverse Lyndon factorization
but $w$ has a unique inverse Lyndon factorization with the border property
(Example \ref{nonunique}, Proposition \ref{UnicaBorder}).
Below we highlight that this unique factorization is the canonical
one defined in \cite{inverseLyndon,inverseLyndonTCS21}.

This special inverse Lyndon factorization
is denoted by $\ICFL$ because it is the counterpart
of the Lyndon factorization $\CFL$ of $w$, when we use (I)inverse words as factors.
Indeed, in \cite{inverseLyndon} it has been proved that $\ICFL(w)$
can be computed in linear time and it is uniquely determined for a word $w$.
See Section \ref{appendice} for definitions of $\ICFL$ and all related notions.
Since $\ICFL(w)$ is the unique inverse Lyndon factorization with the border property,
from now on these two notions will be synonymous.

Below we show another interesting property of $\ICFL$: the last factor of the factorization
is the longest suffix that is an inverse Lyndon word.
Based on this result we provide a new simpler linear algorithm for computing $\ICFL$.

We begin by recalling previously proved results on $\ICFL$, namely Proposition 7.7 in \cite{inverseLyndon}
and Proposition 9.5 in \cite{inverseLyndonTCS21}. They are merged into Proposition \ref{DueRisultati}.

\begin{proposition} \label{DueRisultati}
For any $w \in \Sigma^+$, $\ICFL(w)$ is a grouping of $\CFL_{in}(w)$.
Moreover, $\ICFL(w)$ has the border property.
\end{proposition}

Corollary \ref{ICFLUnicaBorder} is a direct consequence
of Propositions \ref{UnicaBorder}, \ref{BorderCompact}
and \ref{DueRisultati}.

\begin{corollary} \label{ICFLUnicaBorder}
For each $w \in \Sigma^+$,
$\ICFL(w)$ is a compact factorization and it is
is the unique inverse Lyndon factorization of $w$
having the border property.
\end{corollary}

We end the section with a result which has been proved in \cite{IL2023} and which will be used in the
next section.

\begin{proposition} \label{step1}
Let $w \in \Sigma^+$,
let $\CFL_{in}(w) = (\ell_1, \ldots , \ell_h)$ and
let $(\mathcal{C}_{1}, \mathcal{C}_{2}, \ldots , \mathcal{C}_{s})$ be
the decomposition of $\CFL_{in}(w)$ into its non-increasing maximal chains for the prefix order.
Let $w_1, \ldots , w_s$ be words such that
$\CFL_{in}(w_j) = \mathcal{C}_{j}$, $1 \leq j \leq s$.
Then $\ICFL(w)$ is the concatenation of the sequences
$\ICFL(w_1), \ldots , \ICFL(w_s)$, that is,
\begin{equation} \label{EqLocal}
\ICFL(w) = (\ICFL(w_1), \ldots , \ICFL(w_s))
\end{equation}
\end{proposition}

We can now state some results useful to prove the correctness of our algorithm.
First we observe that, thanks to Corollary \ref{ICFLUnicaBorder} and Proposition \ref{step1},
to compute $\ICFL$ we can limit ourselves to the case
in which $\CFL_{in}$ is a chain with respect to the prefix order.

\begin{lemma} \label{step0}
Let $\ell_1, \ldots , \ell_h$ be anti-Lyndon words over $\Sigma$ that form
a non-increasing chain for the prefix order, that is,
$\ell_{1} \geq_p \ell_{2} \geq_p \ldots  \geq_p \ell_{h}$.
If $\ell_1 \not = \ell_2$, then $\ell_{1} \not <_p \ell_2 \cdots \ell_{h}$.
\end{lemma}
\begin{proof}
By contradiction, assume that $\ell_{1}$ is a prefix
of $\ell_2 \cdots \ell_{h}$. Then, $\ell_{1} = \ell_2 \cdots \ell_{t}z$
where either $z = 1$ and $2 < t \leq h$ or $z$ is a nonempty
prefix of $\ell_{t + 1}$, $2 \leq t < h$. Thus either $\ell_{t}$ or $z$ is a nonempty border of
$\ell_{1}$, a contradiction in both cases.
\end{proof}

\begin{remark} \label{twoborders} \cite{inverseLyndonTCS21}
Let $x, y$ two different borders of a same word $w \in \Sigma^+$.
If $x$ is shorter than $y$, then $x$ is a border of $y$.
\end{remark}

\begin{proposition} \label{step2}
Let $w \in \Sigma^+$ and
assume that $\CFL_{in}(w)$ form
a non-increasing chain for the prefix order.
If $(m_1, \ldots , m_k)$ is a factorization of $w$ such that
each $m_j$, $1 \leq j \leq k$, is a concatenation of compact factors in $\CFL_{in}(w)$, then
$(m_1, \ldots , m_k)$ has the border property.
\end{proposition}
\begin{proof}
Let $w \in \Sigma^+$ and
assume that $\CFL_{in}(w)$ form
a non-increasing chain for the prefix order.
Let $(m_1, \ldots , m_k)$ be a factorization of $w$ such that
each $m_j$, $1 \leq j \leq k$, is a concatenation of compact factors in $\CFL_{in}(w)$.
The proof is by induction on $k$. If $k = 1$, then the conclusion follows immediately.
Assume $k > 1$.

Let $w' \in \Sigma^+$ be such that $w = m_1 w'$.
It is clear that $(m_2, \ldots , m_k)$ is a factorization of $w'$ such that
each $m_j$, $2 \leq j \leq k$, is a concatenation of compact factors in $\CFL_{in}(w')$.
Thus, by induction hypothesis, $(m_2, \ldots , m_k)$ has the border property.
It remains to prove that each nonempty border of $m_1$ is not a prefix of $m_{2}$.
The proof is straightforward if $m_1$ is unbordered, thus assume that
$m_1$ is bordered.

Let $\CFL(w) = (\ell_1^{n_1}, \ldots, \ell_r^{n_r})$, where
$\ell_1^{n_1}, \ldots, \ell_r^{n_r}$ are the compact factors in
$\CFL(w)$, that is
$\ell_1, \ldots , \ell_r$ are anti-Lyndon words such that
$\ell_1 \geq_p \ldots \geq_p \ell_h$.
Since $m_i$ is a concatenation of compact factors in $\CFL_{in}(w)$,
there is $h$, $1 \leq h < r$ such that
$$m_1 = \ell_1^{n_1} \cdots \ell_h^{n_h}$$
Notice that $\ell_h$ is a nonempty border of $m_1$. Furthermore,
since $\ell_h$ is unbordered, $\ell_h$ is the shortest nonempty border of $m_1$.

If there were a word $z$ which is a nonempty border
of $m_1$ and also a prefix of $m_{2}$,
by Remark \ref{twoborders}, $\ell_h$ would be a prefix of $m_{2}$.
Therefore, $\ell_h$ would be a prefix
of the word
$\ell_{h + 1}^{n_{h + 1}} \cdots \ell_r^{n_r}$
which contradicts Lemma \ref{step0}.
\end{proof}

\begin{proposition} \label{prop:longest-suffix}
Let $w \in \Sigma^+$ and let
$\ICFL(w) = (m_1, \ldots , m_k)$ be the unique inverse Lyndon factorization of $w$ having the border property.
Then $m_k$ is the longest suffix of $w$ which is an inverse Lyndon word.
\end{proposition}
\begin{proof}
Let $w \in \Sigma^+$ and let $(m_1, \dots, m_k)$ be the unique inverse Lyndon factorization of
$w$ having the border property.
If $k = 1$ we are done. Thus suppose $k > 1$.
By contradiction, suppose that $m_k$ is not the longest suffix
of $w$ that is an inverse Lyndon word. Let $s$ be such longest suffix. Thus,
there exist a nonempty suffix $x$ of $m_j$, $1 \leq j < k$ such that
$s = x m_{j + 1} \cdots m_k$. Furthermore $x$ must be a proper suffix of $m_j$ or we would
have $s = m_j \cdots m_k \ll m_{j + 1} \cdots m_k$ contradicting the hypothesis
that $s$ is inverse Lyndon.

We claim that $x \ll m_{j+1}$.
Indeed, since $m_j$ is an inverse Lyndon word, it holds $x \preceq m_j$.
Thus, if $x \ll m_j$ or $x = m_j$, it immediately follows that $x \ll m_{j+1}$.
Otherwise, $x \leq_p m_j$ and $x$ is a nonempty border of $m_j$.
By Lemma \ref{fundamental} applied to $(m_1, \dots, m_k)$, with $x = \alpha$,
there must exist a prefix $\beta$ of $m_{j+1}$ such that $x \ll \beta$, hence
$x \ll m_{j+1}$.

Since $x \ll m_{j+1}$, we have
$s = x m_{j+1} \cdots m_k \ll m_{j+1} \cdots m_k$, contradicting the hypothesis
that $s$ is an inverse Lyndon word.
\end{proof}

\begin{proposition} \label{step4}
Let $w \in \Sigma^+$ be an inverse Lyndon word, and let $\ell \in \Sigma^+$ be an
anti-Lyndon word. Then:
\begin{enumerate}
\item
If $\ell \ll w$, then for every $k \geq 1$, $\ell^k  w$ is not an inverse Lyndon word.
\item
If $\ell w$ is not an inverse Lyndon word, then $\ell \ll w$.
Furthermore, for every $k \geq 1$, $w$ is the longest suffix of $\ell^k w$ that is an inverse Lyndon word.
\end{enumerate}
\end{proposition}
\begin{proof}
By Lemma \ref{proplexord}, the proof of item 1 is immediate.
Suppose $\ell w$ is not inverse Lyndon. Then, there exists a proper suffix $s$
of $\ell w$ such that $\ell w \preceq s$, hence
$\ell w \ll s$.
Since $\ell$ is anti-Lyndon, for every proper suffix
$x$ of $\ell$ it follows $x \ll \ell$ and consequently
$x w \ll \ell w$.
Thus, $s$ must be a suffix of $w$.
Since $w$ is an inverse Lyndon word, one of the following three cases holds: (1)
$w = s$; (2) $s <_p w$; (3) $s \ll w$.
By $\ell w \ll s$, in each of the three cases it is evident
that $\ell w \ll w$. Thus there are $r, t, t' \in \Sigma^*$ and $a, b \in \Sigma$ with $a < b$ such that
$\ell w = r a t$, $w = r b t'$. If $|\ell| \geq |r a|$, then clearly $\ell \ll w$.
Otherwise, $|\ell| \leq |r |$ and there is $r' \in \Sigma^*$ such that $r = \ell r'$.
Consequently, $w$ starts with $r' a$. On the other hand, $r'$ is a border of $r$, hence
$w = \ell r' b t'$ and $r' b t'$ is a suffix of $w$. This contradicts the fact that
$w$ is an inverse Lyndon word.

For every $k \geq 1$, $w$ is a suffix of $\ell^k w$ that is an inverse Lyndon word.
Let $x$ be a proper nonempty suffix of $\ell$. Of course $x \ll \ell$.
The word $x w$ is not an inverse Lyndon word, otherwise we would have
$\ell \ll w \preceq xw \ll \ell w$, a contradiction.
Moreover, by Lemma \ref{proplexord}, for any $j$, $1 \leq j < k$, we have
$x \ell^j w \ll \ell^j w$ and $x \ell^j w$ is not an inverse Lyndon word.
Finally, by item 1, $\ell^k  w$ is not an inverse Lyndon word.
\end{proof}

\begin{algorithm}
\caption{Compute $\ICFL(w)$,  the unique compact factorization of $w$ having the border property.}\label{algo:factorize}
\begin{algorithmic}[1]
    \Function{Factorize}{$w$}
        \State $(\ell_1^{e_1}, \dots, \ell_n^{e_n}) \gets \Call{CompactFactors}{w}$
            \label{algo:factorize:get-compact}
            \Comment{Compute compact factors of $w$}
        \State $\mathcal{F} \gets \varnothing$
        \State $m' \gets \ell_n^{e_n}$
        \For{$t=n-1 \textbf{ downto } 1$}\label{algo:factorize:loop-start}
            \Comment{Work one compact factor at a time}
            \If{$\ell_t \ll m'$}
                \label{algo:factorize:suffix-condition}
                \Comment{Proposition~\ref{step4}}
                \State $\mathcal{F} \gets (m', \mathcal{F})$
                \State $m' \gets \ell_t^{e_t}$
            \Else
                \State $m' \gets \ell_t^{e_t} \cdot m'$
                \label{algo:factorize:extend}
            \EndIf
        \EndFor\label{algo:factorize:loop-end}
        \State $\mathcal{F} \gets (m', \mathcal{F})$\label{algo:factorize:last-f}
        \State\Return $\mathcal{F}$
    \EndFunction
\end{algorithmic}
\end{algorithm}

We now describe Algorithm~\ref{algo:factorize}. Function $\Call{Factorize}{w}$ will compute
the unique compact factorization of $w$ having the border property.
First, at line~\ref{algo:factorize:get-compact}, it is computed the decomposition of $w$
into its compact factors. Then, the factorization of $w$
is carried out from right to left. Specifically, in accordance with
Proposition~\ref{prop:longest-suffix},
the for-loop at lines~\ref{algo:factorize:loop-start}--\ref{algo:factorize:loop-end}
will search for the longest suffix $m'$ of $w$ that is an inverse Lyndon word.
The update of $m'$ is managed by iteratively applying Proposition~\ref{step4}
at line~\ref{algo:factorize:suffix-condition}. Once such longest suffix is found
(that is, when the condition at line~\ref{algo:factorize:suffix-condition} is true)
it is added to the growing factorization $\mathcal{F}$ and it is initiated a new search
for the longest suffix for the remaining portion of the string.
Otherwise, line~\ref{algo:factorize:extend}, the suffix is extended. In the end, the complete
factorization is returned.

%%%%%%%%%%%%%%%%%%%%%%%%%%%%%%%%%%%%%%%%%%%%%%%%%%
\subsection{Correctness and complexity}

We now prove that Algorithm~\ref{algo:factorize} is correct, that is that it
will compute the unique inverse Lyndon factorization of $w$ having the border property, namely $\ICFL(w)$.
Formally:

\begin{lemma}\label{lem:factorize-correct}
    Let $w \in \Sigma^+$, and let $\mathcal{F}$ be the result of $\Call{Factorize}{w}$.
    Then, $\mathcal{F} = \ICFL(w)$.
\end{lemma}
\begin{proof}
    Let $(\ell_1^{e_1}, \dots, \ell_n^{e_n})$ be the decomposition of $w$ into its
    compact factors, and let $L_t = \ell_t^{e_t} \cdots \ell_n^{e_n}$. We will denote by $m'_t$
    (resp. $\mathcal{F}_t$) the value of $m'$ (resp. $\mathcal{F}$) at the end of iteration
    $t$. We will prove the following loop invariant:
    at the end of iteration $t$, sequence $(m'_t, \mathcal{F}_t)$ is a compact factorization of $L_t$
    having the border property. The claimed result will follow by Corollary~\ref{ICFLUnicaBorder}.
    \begin{description}
        \item[Initialization.] Prior to entering the loop,
            $(m'_n, \mathcal{F}_n) = (\ell_n^{e_n})$ ,
            where the last equality follows from Proposition~\ref{prop:longest-suffix}.
        \item[Maintenance.] Let $t \leq n-1$. By induction hypothesis,
            $\ICFL(L_{t+1}) = (m'_{t+1}, \mathcal{F}_{t+1})$.

            Suppose $\ell_t \ll m'_{t+1}$. Then, by item 1 of Proposition~\ref{step4} $\ell_t \cdot m'_{t+1}$
            is not inverse Lyndon and $m'_{t+1}$ is the longest suffix of $\ell_t^{e_t} \cdot m'_{t+1}$ that is an
            inverse Lyndon word. Thus, by Proposition~\ref{prop:longest-suffix} $m'_{t+1}$ is the last
            factor of any compact factorization of $\ell_t^{e_t} \cdot m'_{t+1}$. Hence,
            $(m'_t, \mathcal{F}_t) = (\ell_t^{e_t}, m'_{t+1}, \mathcal{F}_{t+1})$ is a compact factorization
            of $F_t$ having the border property.

            Now, consider the case where $\ell_t \not\ll m'_{t+1}$. Then, by
            the contrapositive of item 2 of Proposition~\ref{step4},
            $\ell_t \cdot m'_{t+1}$ is inverse Lyndon and thus, again by item 2 of
            Proposition~\ref{step4}, $\ell_t^{e_t} \cdot m'_{t+1}$ is inverse Lyndon.
            Therefore, $(m'_t, \mathcal{F}_t) = (\ell_t^{e_t} \cdot m'_{t+1}, \mathcal{F}_{t+1})$ is a compact factorization having the border property.

        \item[Termination.] After iteration $t=1$, sequence $(m'_1, \mathcal{F}_1) = \ICFL(L_1) = \ICFL(w)$.
    \end{description}
    Finally, line \ref{algo:factorize:last-f} sets $\mathcal{F} = (m'_1, \mathcal{F}_1) = \ICFL(w)$.
\end{proof}

Function $\Call{Factorize}{w}$ has time complexity that is linear in the length
of $w$. Indeed, the sequence of compact factors obtained at line~\ref{algo:factorize:get-compact}
can be computed in linear time in the length
of $w$ by a simple modification of Duval's algorithm (see~\cite{lothaire}).
After that, each iteration $t$ of loop
\ref{algo:factorize:loop-start}--\ref{algo:factorize:loop-end} can be implemented to run
in time $\mathcal{O}(|\ell_t|)$. Indeed, condition $\ell_t \ll m'$ can be checked by naively
comparing $\ell_t$ against $m'$. Furthermore, the update of $m'$ and $\mathcal{F}$ can be done
in constant time: in fact, $\ell_t$, $\ell_t^{e_t}$, $m'$ and $\mathcal{F}$ can all be implemented
as pairs of indexes (in case of the former three) or as a list of indexes (in case of the latter)
of $w$.

%%%%%%%%%%%%%%%%%%%%%%%%%%%%%%%%%%%%%%%%%%%%%%%
\section{Conclusions}

We discover the special connection between the Lyndon factorization under the inverse lexicographic ordering, named $\CFL_{in}$ and the canonical inverse Lyndon factorization, named $\ICFL$: there exists a unique inverse Lyndon factorization having the border property and this unique factorization is $\ICFL$. Moreover each inverse factor of $\ICFL$ is obtained by concatenating compact factors of $\CFL_{in}$. These properties give a constrained structure to $\ICFL$ that deserve to be further explored to characterize properties of words. In particular, we believe the characterization of $\ICFL$ as a compact factorization, proved in the paper, could highlight novel properties related the compression of a word, as investigated in \cite{I2016215}. In particular, the number of compact factors seems to be a measure of repetitiveness of the word to be also used in speeding up suffix sorting of a word.

Finally, we believe that the characterization of $\ICFL$ in terms of $\CFL_{in}$ may be used to extend to $\ICFL$ the \emph{conservation property} proved in \cite{inverseLyndonTCS21} for $\CFL$.
This property shows that the Lyndon factorization of a word $w$ preserves common factors with the factorization of a superstring of $w$. This extends the conservation of Lyndon factors explored for the product $u\cdot v$ of two words $u$ and $v$ \cite{I2016215,apostolico1995fast}.

%%%%%%%%%%%%%%%%%%%%%%%%%%%%%%%%%%%%%%%%%%%%%%%%%%%%%%%%%%%%%%%

\section*{Acknowledgments}
This research was supported by the European Union's Horizon 2020 Research and Innovation Programme under the Marie Sk\l{}odowska-Curie grant agreement PANGAIA No. 872539, by MUR 2022YRB97K, PINC, Pangenome Informatics: from Theory to Applications, and by INdAM-GNCS Project 2023

%Bibliography
\bibliographystyle{unsrt}  
\bibliography{MFCS24Referenze}  

%%%%%%%%%%%%%%%%%%%%%%%%%%%%%%%%%%%%%%%%%%%%%%%%%%%%%%%%%%%%%%%%%%%

\appendix

\section{The canonical inverse Lyndon factorization} \label{appendice}

In this section we summarize the relevant material on
the {\it canonical inverse Lyndon factorization} and
we refer to \cite{inverseLyndon,inverseLyndonTCS21}
for a thorough discussion on this topic.

If $w$ is an inverse Lyndon word, then $\ICFL(w) = w$.
Otherwise, $\ICFL(w)$ is recursively defined.
The first factor of $\ICFL(w)$
is obtained by a special pair $(p , \overline{p})$ of words, named
the canonical pair associated with $w$, which in turn is obtained by the shortest nonempty
prefix $z$ of $w$ such that $z$ is not an inverse Lyndon word.
Proposition 6.2 in \cite{inverseLyndonTCS21} provides the following characterization of the
pair $(p , \overline{p})$.

\begin{proposition} \label{characterization}
Let $w \in \Sigma^+$ be a word which is not an inverse Lyndon word.
A pair of words $(p , \overline{p})$ is the canonical pair associated with $w$
if and only the following conditions are satisfied.
\begin{itemize}
\item[(1)]
$z = p \overline{p}$ is the shortest nonempty prefix of $w$ which is not an inverse Lyndon word.
\item[(2)]
$p = ras$ and $\overline{p} = rb$, where $r,s \in \Sigma^*$, $a,b \in \Sigma$ and
$r$ is the shortest prefix of $p \overline{p}$ such that $p \overline{p} = rasrb$,
with $a < b$.
\item[(3)]
$\overline{p}$ is an inverse Lyndon word.
\end{itemize}
\end{proposition}

Given a word $w$ which is not an inverse Lyndon word,
Proposition \ref{characterization} suggests a method to identify
the canonical pair $(p , \overline{p})$ associated with $w$:
just find the shortest nonempty prefix $z$ of $w$ which is not an inverse Lyndon word
and then a factorization $z = p \overline{p}$ such that conditions (2) and (3) in Proposition
\ref{characterization} are satisfied.

The canonical inverse Lyndon factorization has been also recursively defined.

\begin{definition} \label{def:ICFL}
Let $w \in \Sigma^+$. \\
{\rm (Basis Step)}
If $w$ is an inverse Lyndon word,
then $\ICFL(w) = (w)$. \\
{\rm (Recursive Step)}
If $w$ is not an inverse Lyndon word,
let $(p,\overline{p})$ be the canonical pair associated with $w$ and let
$v \in \Sigma^*$ such that $w = pv$.
Let $\ICFL(v) = (m'_1, \ldots, m'_{k})$ and let
$r,s \in \Sigma^*$, $a,b \in \Sigma$ such that $p = ras$, $\overline{p} = rb$ with $a < b$.
$$\ICFL(w) = \begin{cases} (p, \ICFL(v)) & \mbox{ if } \overline{p} = rb \leq_{p} m'_1 \\
(p m'_1, m'_2, \ldots, m'_{k}) & \mbox{ if } m'_1 \leq_{p} r \end{cases}$$
\end{definition}

The following example is in \cite{inverseLyndon}.

\begin{example} \label{nonunique2}
{\rm Let $\Sigma = \{a,b,c,d \}$ with $a < b < c < d$,
$w = dabadabdabdadac$. We have
$\CFL_{in}(w) = (daba, dab, dab, dadac)$ and
$\ICFL(w) = (daba, dabdab, dadac)$.
Consider
$z = dabdadacddbdc$.
We have $\ICFL(z) = \CFL_{in}(z) = (dab,dadac,ddbdc)$.}
\end{example}

\end{document}